\documentclass{article}

\usepackage{amsmath}
\usepackage{amssymb}
\usepackage{amsthm}
\usepackage{array}
\usepackage{subfigure}
\usepackage{epsfig}
\usepackage{latexsym}
\usepackage{algorithm}
\usepackage{algorithmic}
\usepackage[latin1]{inputenc}
\usepackage{graphicx}

\newtheorem{theorem}{Theorem}
\newtheorem{lemma}{Lemma}

\newcommand{\avg}[1]{\mathop{\mathrm{avg} \{f(y)\}}_{y \in #1}}

\newcommand{\avgf}[2]{\mathop{\mathrm{avg} \{#2(y)\}}_{y \in #1}}

\title{Elementary Components of the Quadratic Assignment Problem}
\author{Francisco Chicano, Gabriel Luque and Enrique Alba}
\date{}

\begin{document}

\clubpenalty=10000
\widowpenalty=10000

\maketitle

\begin{abstract}
The Quadratic Assignment Problem (QAP) is a well-known NP-hard
combinatorial optimization problem that is at the core of many
real-world optimization problems. We prove that QAP can be written as the sum of
three elementary landscapes when the swap neighborhood is used. We
present a closed formula for each of the three elementary components
and we compute bounds for the autocorrelation coefficient.
\end{abstract}

\section{Introduction}


We will define a \emph{landscape} for a combinatorial problem using
a triple $(X,N,f)$, where
$f : X \mapsto \mathbb{R}$ defines the objective function
and the \emph{neighborhood operator} $N$ assigns
a set of neighboring solutions $N(x) \in X$ to each solution $x$.
 If $y \in N(x)$
then $y$ is a neighbor of $x$.
The landscape that is induced can be used as a search
space for optimization using local search.

There is a special kind of landscape which is of particular interest
due to their properties. They are the {\em elementary landscapes},
and are characterized by the following equation:

\begin{equation}
\label{eqn:grover} \avg{N(x)} = f(x) + \frac{k}{d} \left( \bar{f} -
f(x) \right)
\end{equation}
where $d$ is the size of the neighborhood, $|N(x)|$, which we assume
the same for all the solutions in the search space, $\bar{f}$ is the
average solution evaluation over the entire search space, and $k$ is
a characteristic constant. Equation (\ref{eqn:grover}) is usually called
\emph{Grover's wave equation} and makes
it possible to compute the average value of the fitness function $f$
evaluated over all of the neighbors of $x$; we denote this average
using $\avg{N(x)}$:
\[ \avg{N(x)} = \frac{1}{|N(x)|} \sum_{y \in N(x)} f(y) \]
Other properties also follow.
Assuming $f(x) \neq \bar{f}$ then
\[f(x) < \min \left\{ \avg{N(x)}, \bar{f} \right\} \vee  f(x) > \max \left\{\avg{N(x)} ,\bar{f}\right\}. \]
\noindent This implies that all maxima are greater than $\bar{f}$
and all minima are less than $\bar{f}$ \cite{Whitley2008}.
%
%
A landscape $(X,N,f)$ is not always
elementary, but even in this case it is
possible to characterize the function $f$ as the sum of elementary
landscapes~\cite{Sutton2009}, called \emph{elementary components} of
the landscape.

The Quadratic Assignment Problem (QAP) is an NP-hard combinatorial
optimization problem~\cite{Garey:Johnson1979}. A lot of research has
been devoted to analyze and solve the QAP. Some other problems can
be formulated as special cases of the QAP. One important example is
the Traveling Salesman Problem (TSP). The QAP is not an elementary
landscape when the swap neighborhood is
considered~\cite{Angel:Zissimopoulos2000b}. But, to the best of our
knowledge the exact expressions for the elementary components of the
QAP are not known.

Such decomposition could be useful from the theoretical and
practical points of view. In theory, the landscape decomposition of
QAP can be used to compute the exact expression of the
autocorrelation functions and the autocorrelation
coefficient~\cite{Angel:Zissimopoulos2000b}. In practice, the
landscape decomposition together with the Grover's wave equation can
be used to compute the average value of the objective function in
the neighborhood, which can be used as a base for new operators or
algorithms. In particular, a new family of selection operators can
be designed which select the individuals according to the average
fitness value in the neighborhood of a solution $x$ instead of using
the fitness value of the solution itself. These selection operators
could be especially useful to distinguish solutions that are in
plateaus.

We present here the elementary landscape decomposition of QAP. In the next section we present the formal
definition of QAP. Section~\ref{sec:decomposition} presents the main
result and its proof. Finally, we conclude in Section~\ref{sec:conclusions}
with some conclusions and future work.

\section{Quadratic Assignment Problem}
\label{sec:problem}

Let $P$ be a set of $n$ facilities and $L$ a set of $n$ locations.
For each pair of locations $i$ and $j$, an arbitrary distance is specified
$r_{ij}$ and for each pair of facilities $p$ and $q$, a flow is
specified $w_{pq}$. The Quadratic Assignment Problem (QAP) consists
in assigning the facilities of $P$ to the locations in $L$ in such a
way that the total cost of the assignment is minimized. Each
location can only contain one facility. For each pair of facilities
the cost is computed as the product of the weight associated to the
facilities and the distance between the locations in which the
facilities are. The total cost is the sum of all the costs
associated to each pair of facilities. One solution to this problem
is a bijection between $P$ and $L$, that is, $x: P \rightarrow L$ such
that $x$ is bijective. Without loss of generality we can just assume
that $P=L=\{1,2,\ldots,n\}$ and each solution $x$ is a permutation
in $S_n$, the set permutations of $\{1,2,\ldots,n\}$. The cost
function to be minimized can be formally defined as:

\begin{equation}
\label{eqn:fitness} f(x) =\sum_{i,j=1}^{n} {r_{ij} w_{x(i)x(j)}}
\end{equation}

\section{Decomposition of QAP}
\label{sec:decomposition}

In the previous section we defined the search space and the
objective function. In order to completely define a landscape we
need to define the neighborhood. The neighborhood $N$ considered
here is the swap or 2-exchange neighborhood, in which two solutions
are neighboring if one can be obtained from the other one by a swap
(exchange of two elements) in the permutation. In this section we
prove that the QAP is composed at most by three elementary
components and we give an expression for them.


Let us start rewriting (\ref{eqn:fitness}). In order to analyze the
elementary components of the fitness functions associated to a
problem class, like QAP, it is useful to separate in the formal
definition of the objective function the information that is
particular of a given instance (the data of the instance) from the
general issues that characterize the class of the problem. In the
case of QAP, the information related to the particular instance is
included in the distance matrix $(r_{ij})$ and the weight matrix
$(w_{pq})$. The question now is: how to separate the data of the instance in the fitness
formulation. There are different ways to do
it, but we are interested in linear combinations of functions where
the coefficients of the functions are associated to the particular
instances. The reason for this is that any linear combination of
elementary functions (with the same
characteristic constant $k$) is also an elementary function.
With this idea in mind, it is not
difficult to see that Equation (\ref{eqn:fitness}) can be written
using the following linear combination:

\begin{equation}
\label{eqn:fitness2} f(x) = \sum_{i,j=1}^{n} \sum_{p,q=1}^{n} r_{i
j} w_{p q} \delta_{x(i)}^{p} \delta_{x(j)}^{q}
\end{equation}
where we used the Kronecker's delta. The problem-related part of the
fitness function is, thus, the product
$\delta_{x(i)}^{p}\delta_{x(j)}^{q}$. At this point we can go
further and deal with a more general objective function. In
(\ref{eqn:fitness2}) the value of the product $r_{ij} w_{pq}$
depends on $i$, $j$, $p$, and $q$ in a particular way, but is not
the most general one. Using multilinear algebra concepts, the
previous product is a four-rank tensor that has been computed as a
tensor product of two two-rank tensors (matrices), which is a
special case of four-rank tensor. In the most general case we can
define a four-rank tensor to replace the product. Let us call
$\psi_{ijpq}$ the new general four-rank tensor and let us define the
parameterized function $\varphi_{(i,j),(p,q)} (x) =
\delta_{x(i)}^{p}\delta_{x(j)}^{q}$. Then we can rewrite the fitness
function as:
\begin{equation}
\label{eqn:fitness3} f = \sum_{i,j,p,q=1}^{n} \psi_{ijpq}
\varphi_{(i,j),(p,q)}
\end{equation}
and we can focus our analysis on $\varphi_{(i,j),(p,q)}$, since any
result on it can be extended to any linear combination of $\varphi$
functions, and, thus, to $f$. Now, the objective function of the QAP
is just a particular case of our new objective function $f$, in
which $\psi_{ijpq} = r_{ij} w_{pq}$.

If $i=j$ and $p=q$ in the functions $\varphi_{(i,j),(p,q)}$, then we
have $\varphi_{(i,i),(p,p)}(x)=\delta_{x(i)}^p$. For this particular
case we have the following

\begin{lemma}
Considering the swap neighborhood the function
$\varphi_{(i,i),(p,p)}$ is an elementary landscape with $k=n$.
\end{lemma}
\begin{proof}
In the following, for the sake of clarity we will remove all the
parameters from the name of the function when there is no confusion.
The function $\varphi$ is elementary if and only if there exist two
constants $a$ and $b$ such that the following expression holds for
all the solutions:
\[ \avgf{N(x)}{\varphi} = a \varphi(x) + b\]
In order to reduce the expressions we multiply the previous
expression by the size of the neighborhood, which is
$d=\frac{n(n-1)}{2}$. We then obtain:
\begin{equation}
\sum_{y\in N(x)}\varphi(y) = c \varphi(x) + e \label{eqn:sum1}
\end{equation}
where $c=ad$ and $e=bd$. Now, we compute the exact expression of
$\sum_{y\in N(x)}\varphi(y)$ for the two different values that
$\varphi$ can take:
\begin{itemize}
\item Case $\varphi(x)=1$ (in this case $x(i)=p$). From the neighboring solutions there are $n-1$ with $\varphi(y)=0$
and the remaining neighbors have a value $\varphi(y)=1$. Then we can
write:
\[ \sum_{y\in N(x)}\varphi(y) = (d-n+1)\]

\item Case $\varphi(x)=0$ (in this case $x(i)\neq p$). From the neighboring solutions there is only one with
$\varphi(y)=1$. The remaining neighbors have a value $\varphi(y)=0$.
Then we can write:
\[ \sum_{y\in N(x)}\varphi(y) = 1\]
\end{itemize}

Now we use Equation (\ref{eqn:sum1}) to obtain the following linear
equation system:
\begin{equation}
\nonumber \left( \begin{array}{cc}
1 & 1 \\
0 & 1 \\
\end{array}
\right) \left(
\begin{array}{c} c \\ e
\end{array}
\right) = \left( \begin{array}{c}
d-n+1\\
1\\
\end{array}
\right)
\end{equation}

The solution of the previous system is $c=d-n$ and $e=1$; so we have
$a=1-n/d$ and $b=1/d$. Then, we can write
\begin{equation}
\avgf{N(x)}{\varphi} = \left(1-\frac{n}{d}\right) \varphi(x) + \frac{1}{d} = \varphi(x) + \frac{n}{d} \left(\frac{1}{n} - \varphi(x)\right) \\
\end{equation}
and we conclude that $\varphi_{(i,i),(p,p)}$ is an elementary
landscape with $k=n$ and average $\bar{\varphi}_{(i,i),(p,p)}=1/n$.
\end{proof}

Before proving the main result of this section we need to introduce
a family of auxiliary functions that map permutations to
$\mathbb{R}$:

\begin{equation}
\label{eqn:phi}
\phi_{(i,j),(p,q)}^{\alpha,\beta,\gamma,\varepsilon,\zeta}(x) = \left\{
\begin{array}{ll}
\alpha & \mbox{if $x(i)=p \wedge x(j)=q$} \\
\beta & \mbox{if $x(i)=q \wedge x(j)=p$} \\
\gamma & \mbox{if $x(i)=p \oplus x(j)= q$} \\
\varepsilon & \mbox{if $x(i)=q \oplus x(j)= p $} \\
\zeta & \mbox{if $x(i) \neq p,q \wedge x(j) \neq p,q$} \\
\end{array}
\right.
\end{equation}
where $1\leq i,j,p,q \leq n$ are integer values with $i \neq j$ and
$p\neq q$ and $\alpha, \beta, \gamma, \varepsilon, \zeta \in
\mathbb{R}$. All the previous values are parameters of the family of
functions. We denote with $\oplus$ the exclusive-or operator.
The previous functions are valuable thanks to the
following

\begin{lemma}
Considering the swap neighborhood the function
$\phi_{(i,j),(p,q)}^{\alpha,\beta,\gamma,\varepsilon,\zeta}$ is an
elementary landscape in the following cases:
\begin{itemize}
\item $\alpha=n-3$, $\beta=1-n$, $\gamma=-2$, $\varepsilon=0$, $\zeta=-1$ with $k=2n$
\item $\alpha=n-3$, $\beta=n-3$, $\gamma=0$, $\varepsilon=0$, $\zeta=1$ with $k=2(n-1)$
\item $\alpha=2n-3$, $\beta=1$, $\gamma=n-2$, $\varepsilon=0$, $\zeta=-1$ with $k=n$
\end{itemize}
\end{lemma}
\begin{proof}
In the following, for the sake of clarity we will remove
all the parameters from the name of the function when there
is no confusion.
The function $\phi$ is elementary if and only if there exist two constants $a$ and $b$ such that the following expression holds for all the solutions:
\[ \avgf{N(x)}{\phi} = a \phi(x) + b\]
In order to reduce the expressions we multiply the previous expression by the size of the neighborhood, which is $d=\frac{n(n-1)}{2}$. We then obtain:
\begin{equation}
\sum_{y\in N(x)}\phi(y) = c
\phi(x) + e \label{eqn:sum}
\end{equation}
where $c=ad$ and $e=bd$.

\begin{figure}[!ht]
\centering
\includegraphics[width=6cm]{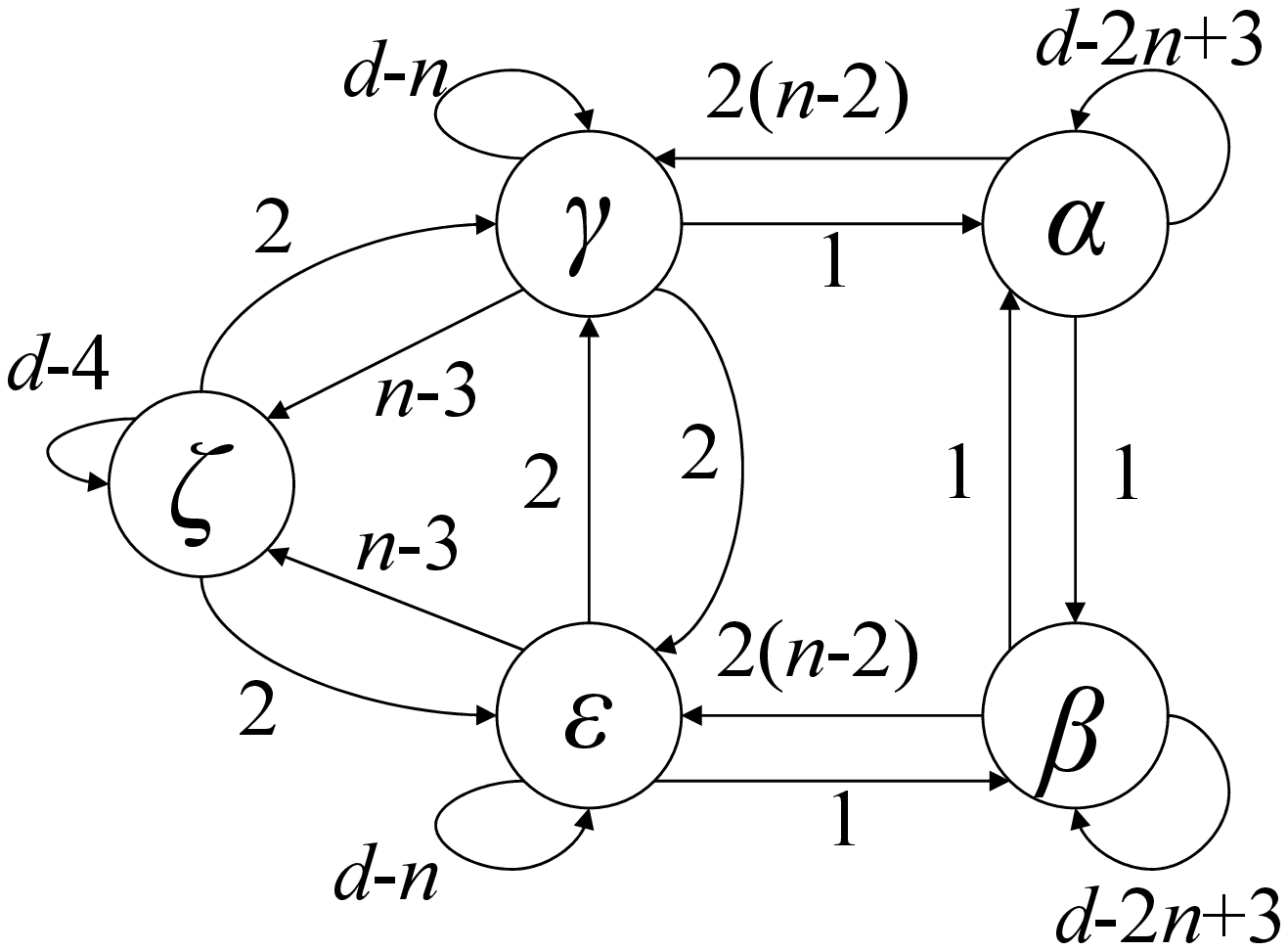}
\caption{Transition graph for functions
$\phi_{(i,j),(p,q)}^{\alpha,\beta,\gamma,\varepsilon,\zeta}$.}
\label{fig:phi-graph} \vspace{-12pt}
\end{figure}

We distinguish five different cases which are symbolically
represented in Figure~\ref{fig:phi-graph}. In the figure, each node
represents the set of solutions for which one of the five branches
in (\ref{eqn:phi}) is true. We label the nodes with the value that
$\phi$ takes for all the solutions in that node. There exists an arc
$(i,j)$ if all the solutions in node $i$ have at least one
neighboring solution in node $j$. The label of arc $(i,j)$ is the
number of neighbors that any solution in $i$ has in $j$ . Now, we
compute the exact expression of $\sum_{y\in N(x)}\phi(y)$ for the
five different values that $\phi$ can take:

\begin{itemize}
\item Case $\phi(x)=\alpha$. In this case $x(i)=p$ and $x(j)=q$. From the neighboring solutions there is one with $\phi(y)=\beta$
and $2(n-2)$ solutions with $\phi(y)=\gamma$. The remaining neighbors have a value $\phi(y)=\alpha$. Then we can write:
\[ \sum_{y\in N(x)}\phi(y) = \beta + 2(n-2) \gamma + (d-2n+3) \alpha\]

\item Case $\phi(x)=\beta$. In this case $x(i)=q$ and $x(j)=p$. From the neighboring solutions there is one with $\phi(y)=\alpha$
and $2(n-2)$ solutions with $\phi(y)=\varepsilon$. The remaining
neighbors have a value $\phi(y)=\beta$. Then we can write:
\[ \sum_{y\in N(x)}\phi(y) = \alpha + 2(n-2) \varepsilon + (d-2n+3) \beta\]

\item Case $\phi(x)=\gamma$. In this case $x(i)=p$ or $x(j)= q$, but not both. From the neighboring solutions there is one with $\phi(y)=\alpha$,
two neighbors with $\phi(y)=\varepsilon$, and $n-3$ neighbors with
$\phi(y)=\zeta$. The remaining neighbors have a value
$\phi(y)=\gamma$. Then we can write:
\[ \sum_{y\in N(x)}\phi(y) = \alpha + 2 \varepsilon + (n-3) \zeta + (d-n) \gamma\]

\item Case $\phi(x)=\varepsilon$. In this case $x(i)=q$ or $x(j)= p$, but not both. From the neighboring solutions there is one with $\phi(y)=\beta$,
two neighbors with $\phi(y)=\gamma$, and $n-3$ neighbors with
$\phi(y)=\zeta$. The remaining neighbors have a value
$\phi(y)=\varepsilon$. Then we can write:
\[ \sum_{y\in N(x)}\phi(y) = \beta + 2 \gamma + (n-3) \zeta + (d-n) \varepsilon\]

\item Case $\phi(x)=\zeta$. In this case $x(i) \neq p,q$ and $x(j) \neq p,q$. From the neighboring solutions there are two with $\phi(y)=\gamma$
and two neighbors with $\phi(y)=\varepsilon$. The remaining neighbors
have a value $\phi(y)=\zeta$. Then we can write:
\[ \sum_{y\in N(x)}\phi(y) = 2 \gamma + 2 \varepsilon + (d-4) \zeta\]
\end{itemize}

Now we use Equation (\ref{eqn:sum}) to obtain the following linear
equation system:

\begin{equation}
\nonumber
\left( \begin{array}{rr}
\alpha & 1 \\
\beta & 1 \\
\gamma & 1 \\
\varepsilon & 1 \\
\zeta & 1 \\
\end{array}
\right)
\left(
\begin{array}{r} c \\ e
\end{array}
\right) =
\left( \begin{array}{r}
\beta + 2(n-2) \gamma + (d-2n+3) \alpha\\
\alpha + 2(n-2) \varepsilon + (d-2n+3) \beta \\
\alpha + 2 \varepsilon + (n-3) \zeta + (d-n) \gamma \\
\beta + 2 \gamma + (n-3) \zeta + (d-n) \varepsilon \\
2 \gamma + 2 \varepsilon + (d-4) \zeta \\
\end{array}
\right)
\end{equation}

The previous system has five equations and two variables, $c$ and
$e$, so it could be unsolvable. However, the system can be solved
for some value combinations of $\alpha$, $\beta$, $\gamma$,
$\varepsilon$, $\zeta$. In particular, the system can be solved for the
value combinations mentioned in the statement, that is:

\begin{enumerate}
\item $\alpha=n-3$, $\beta=1-n$, $\gamma=-2$, $\varepsilon=0$, $\zeta=-1$
\item $\alpha=n-3$, $\beta=n-3$, $\gamma=0$, $\varepsilon=0$, $\zeta=1$
\item $\alpha=2n-3$, $\beta=1$, $\gamma=n-2$, $\varepsilon=0$, $\zeta=-1$
\end{enumerate}

This does not mean that these are the only combinations of parameter values
for which the system can be solved. They are just three combinations of special
interest for the goal of this section.
It should be noticed here that the linear system does not depend on the values of
$i$, $j$, $p$, and $q$. Thus, the solutions to the system are also independent of the
values of the mentioned parameters.

Let us study the values of $a, b, c, e$ for the first parameter
combination, that is, $\alpha=n-3$, $\beta=1-n$, $\gamma=-2$,
$\varepsilon=0$, and $\zeta=-1$. The solution of the linear system is
$c=\frac{n(n-5)}{2}$ and $e=-2n$, and, thus: $a=(1-2n/d)$ and
$b=-2n/d$. In order to simplify the notation, let us define
$\Omega_{(i,j),(p,q)}^{1} = \phi_{(i,j),(p,q)}^{n-3,1-n,-2,0,-1}$.
Then, we can write

\begin{eqnarray}
\nonumber & & \avgf{N(x)}{\Omega_{(i,j),(p,q)}^{1}} = \left(1-\frac{2n}{d}\right) \Omega_{(i,j),(p,q)}^{1}(x) - \frac{2n}{d} =\\
& & =\Omega_{(i,j),(p,q)}^{1}(x) +
\frac{2n}{d}\left(-1-\Omega_{(i,j),(p,q)}^{1}(x)\right)
\end{eqnarray}
and we conclude that $\Omega_{(i,j),(p,q)}^{1}$ is an
elementary landscape with $k=2n$ and average
$\bar{\Omega}^{1}_{(i,j),(p,q)}=-1$.

Let us now focus on the second parameter combination, that is,
$\alpha=\beta=n-3$, $\gamma=\varepsilon=0$, and $\zeta=1$. The solution
of the linear system is $c=\frac{(n-1)(n-4)}{2}$ and $e=2(n-3)$,
and, thus: $a=1-2(n-1)/d$ and $b=2(n-3)/d$. Introducing the notation
$\Omega_{(i,j),(p,q)}^{2} = \phi_{(i,j),(p,q)}^{n-3,n-3,0,0,1}$ we
can write

\begin{eqnarray}
 & & \avgf{N(x)}{\Omega_{(i,j),(p,q)}^{2}} =\\
\nonumber & & = (1-2(n-1)/d) \Omega_{(i,j),(p,q)}^{2}(x) + 2(n-3)/d =\\
& & =\Omega_{(i,j),(p,q)}^{2}(x) +
\nonumber\frac{2(n-1)}{d}\left(\frac{n-3}{n-1}-\Omega_{(i,j),(p,q)}^{2}(x)\right)
\end{eqnarray}
and we conclude that $\Omega_{(i,j),(p,q)}^{2}$ is an
elementary landscape with $k=2(n-1)$ and
$\bar{\Omega}_{(i,j),(p,q)}^{2}=(n-3)/(n-1)$.

Finally, let us analyze the third parameter combination, that is,
$\alpha=2n-3$, $\beta=1$, $\gamma=n-2$, $\varepsilon=0$ and $\zeta=-1$.
The solution of the linear system is $c=\frac{n(n-3)}{2}$ and $e=n$,
and, thus: $a=1-n/d$ and $b=n/d$. Introducing the notation
$\Omega_{(i,j),(p,q)}^{3} = \phi_{(i,j),(p,q)}^{2n-3,1,n-2,0,-1}$ we
can write

\begin{eqnarray}
\nonumber & & \avgf{N(x)}{\Omega_{(i,j),(p,q)}^{3}} = \left(1-\frac{n}{d}\right) \Omega_{(i,j),(p,q)}^{3}(x) + \frac{n}{d} =\\
& & = \Omega_{(i,j),(p,q)}^{3}(x) +
\frac{n}{d}\left(1-\Omega_{(i,j),(p,q)}^{3}(x)\right)
\end{eqnarray}
and we conclude that $\Omega_{(i,j),(p,q)}^{3}$ is an elementary
landscape with $k=n$ and $\bar{\Omega}_{(i,j),(p,q)}^{3}=1$.
\end{proof}

Now, we are in conditions of presenting the main result of the paper, which is
the following

\begin{theorem}
\label{thm:decomposition} For the swap neighborhood defined above,
the function $f$ defined in (\ref{eqn:fitness3}) is the sum of at
most three elementary landscapes with constants $k_1=2n$,
$k_2=2(n-1)$, and $k_3=n$.
\end{theorem}
\begin{proof}
The functions $\varphi_{(i,j),(p,q)}$ defined above can be written
using the auxiliary functions $\Omega_{(i,j),(p,q)}^{1}$,
$\Omega_{(i,j),(p,q)}^{2}$, and $\Omega_{(i,j),(p,q)}^{3}$ when
$i\neq j$ and $p\neq q$ as (we omit the subindices for clarity)

\begin{equation}
\varphi = \frac{1}{2n} \Omega^1 + \frac{1}{2(n-2)} \Omega^2 + \frac{1}{n(n-2)} \Omega^3
\end{equation}
This can be easily appreciated with the help of Table~\ref{tab:components}.

\begin{table}[!h]
\begin{center}
\begin{tabular}{|l||r|r|r||r|}
\hline
Condition & $\frac{\Omega^1}{2n}$ & $\frac{\Omega^2}{2(n-2)}$ & $\frac{\Omega^3}{n(n-2)}$ &  $\sum$ \\
\hline
$x(i)=p \wedge x(j)=q$               & $\frac{n-3}{2n}$ & $\frac{n-3}{2(n-2)}$ & $\frac{2n-3}{n(n-2)}$ & 1 \\
$x(i)=q \wedge x(j)=p$               & $\frac{1-n}{2n}$ & $\frac{n-3}{2(n-2)}$ & $\frac{1}{n(n-2)}$    & 0 \\
$x(i)=p \oplus x(j)=q$           & $\frac{-2}{2n}$  & $0$                  & $\frac{n-2}{n(n-2)}$  & 0 \\
$x(i)=q \oplus x(j)=p $          & $0$              & $0$                  & $0$                   & 0 \\
$x(i) \neq p,q \wedge x(j) \neq p,q$ & $\frac{-1}{2n}$  & $\frac{1}{2(n-2)}$   & $\frac{-1}{n(n-2)}$   & 0 \\
\hline
\end{tabular}
\end{center}
\caption{Elementary components and their sum.}
\label{tab:components}
\end{table}

Since the $\Omega$ family of functions are elementary, the $\varphi$
family of functions are a sum of three elementary components,
namely: $\frac{1}{2n} \Omega^1$, $\frac{1}{2(n-2)} \Omega^2$, and
$\frac{1}{n(n-2)} \Omega^3$. This decomposition of $\varphi$ allows
us to write the fitness function $f$ as a decomposition of
elementary landscapes in the following way:

\begin{align*}
f &= \!\!\!\!\!\! \sum_{\scriptsize\begin{array}{c}i,j,p,q=1\\i\neq
j\\ p\neq q\end{array}}^{n} \!\!\!\!\!\!\!\! \psi_{ijpq}
~\varphi_{(i,j),(p,q)} +
\sum_{i,p=1}^{n} \psi_{iipp} ~\varphi_{(i,i),(p,p)} = \\
& = \!\!\!\!\!\!\!\!\!\!\!\!
\sum_{\scriptsize\begin{array}{c}i,j,p,q=1\\i\neq j\\ p\neq
q\end{array}}^{n} \!\!\!\!\!\!\!\!\!\!\! \psi_{ijpq}~
\left(\frac{\Omega_{(i,j),(p,q)}^1}{2n} + \frac{\Omega_{(i,j),(p,q)}^2}{2(n-2)}  + \frac{\Omega_{(i,j),(p,q)}^3}{n(n-2)}\right) +\\
&+\sum_{i,p=1}^{n} \psi_{iipp} ~\varphi_{(i,i),(p,p)}
\end{align*}

The elementary components of $f$ are:
\begin{eqnarray}
\label{eqn:fc1}
&&\hspace{-40pt}f_{c1} = \sum_{\scriptsize\begin{array}{c}i,j,p,q=1\\i\neq j\\ p\neq q\end{array}}^{n} \!\!\!\!\!\!\!\!\!\!\! \psi_{ijpq}~ \frac{\Omega_{(i,j),(p,q)}^1}{2n} \\
\label{eqn:fc2}
&&\hspace{-40pt}f_{c2} = \sum_{\scriptsize\begin{array}{c}i,j,p,q=1\\i\neq j\\ p\neq q\end{array}}^{n} \!\!\!\!\!\!\!\!\!\!\! \psi_{ijpq}~  \frac{\Omega_{(i,j),(p,q)}^2}{2(n-2)} \\
\label{eqn:fc3} &&\hspace{-40pt}f_{c3} = \!\!\!\!
\sum_{\scriptsize\begin{array}{c}i,j,p,q=1\\i\neq j\\ p\neq
q\end{array}}^{n} \!\!\!\!\!\!\!\!\!\!\! \psi_{ijpq}~
\frac{\Omega_{(i,j),(p,q)}^3}{n(n-2)} + \sum_{i,p=1}^{n} \psi_{iipp}
~\varphi_{(i,i),(p,p)}
\end{eqnarray}
where the functions $f_{c1}$, $f_{c2}$, and $f_{c3}$ are elementary
with constants $k_1=2n$, $k_2=2(n-1)$ and $k_3=n$, respectively,
because they are a linear combination of elementary functions. Thus,
$f$ can be written in a compact form as $f = f_{c1} + f_{c2} +
f_{c3}$.
\end{proof}

In the statement of the theorem we say that the number of elementary
components is three $\emph{at most}$. That is, this number cannot be
larger than three, but it could be lower.
It is possible that for some particular instances the number of
elementary landscapes could be reduced (we will see later that this
happens for the TSP).


Unlike elementary landscapes, when the fitness
function is the sum of several elementary components with different
characteristic constants, the average value in the neighborhood does
not linearly depend on the fitness function. Since, $f_{c1}$,
$f_{c2}$, and $f_{c3}$ are elementary components, the Grover's wave
equation (\ref{eqn:grover}) can be applied to them. And we can
compute the average value in the neighborhood in the following way:

\vspace{-12pt}
\begin{align*}
\avgf{N(x)}{f} &= \avgf{N(x)}{f_{c1}} + \avgf{N(x)}{f_{c2}} + \avgf{N(x)}{f_{c3}} \\
&= f_{c1}(x) + f_{c2}(x) + f_{c3}(x) + \frac{k_1}{d} \left( \bar{f}_{c1} - f_{c1}(x) \right) +\\
&+ \frac{k_2}{d} \left( \bar{f}_{c2} - f_{c2}(x) \right) + \frac{k_3}{d} \left( \bar{f}_{c3} - f_{c3}(x) \right) = \\
&= f(x) + \frac{4}{n-1} \left( \bar{f}_{c1} - f_{c1}(x) \right) + \frac{4}{n} \left( \bar{f}_{c2} - f_{c2}(x) \right) \\
&+ \frac{2}{n-1} \left( \bar{f}_{c3} - f_{c3}(x) \right)
\end{align*}

\section{Conclusions and Future Work}
\label{sec:conclusions}

We have proven that QAP can be decomposed as a sum of three
elementary landscapes when the swap neighborhood is used
(Theorem~\ref{thm:decomposition}). We have presented the exact
expressions for the three components in Equations (\ref{eqn:fc1}),
(\ref{eqn:fc2}), and (\ref{eqn:fc3}). The elementary components
of QAP allow one to compute the average value of the objective function
in the neighborhood of a given solution $x$ using the evaluation of
the three components in $x$.

\section{Acknowledgments}

The authors thank Professor L. Darrell Whitley for his useful
discussions during his stay at the University of Málaga. This
research has been partially funded by the Spanish Ministry of
Science and Innovation and FEDER under contract TIN2008-06491-C04-01
(the M$^*$ project) and the Andalusian Government under contract
P07-TIC-03044 (DIRICOM project).


\end{document}